\documentclass[conference]{IEEEtran}
\IEEEoverridecommandlockouts

\usepackage{mathpazo}
\usepackage{times}
\usepackage{cite}
\usepackage{amsmath,amssymb,amsfonts}
\usepackage{algorithmic}
\usepackage{graphicx}
\usepackage{textcomp}
\usepackage{xcolor}
\usepackage{breqn}
\usepackage{mathtools}
\usepackage{bbm}
\usepackage{bm}
\usepackage{amsthm}
\usepackage{balance}

\DeclareSymbolFont{cmsymbols}{OMS}{cmsy}{m}{n}
\let\emptyset\relax
\DeclareMathSymbol{\emptyset}{\mathord}{cmsymbols}{"3B}

\def\BibTeX{{\rm B\kern-.05em{\sc i\kern-.025em b}\kern-.08em
    T\kern-.1667em\lower.7ex\hbox{E}\kern-.125emX}}

\newtheorem{theorem}{Theorem}

\newtheorem{remark}{Remark}

\begin{document}

\pagestyle{plain}

\title{Private Contiguous-Block Retrieval}

\author{
\IEEEauthorblockN{Maha Issa and Anoosheh Heidarzadeh}
\IEEEauthorblockA{
Department of Electrical and Computer Engineering\\
Santa Clara University, Santa Clara, CA, USA\\
\{missa,aheidarzadeh\}@scu.edu
}
}

\maketitle

\thispagestyle{plain}

\begin{abstract}
We introduce the \emph{Private Contiguous-Block Retrieval (PCBR)} problem, where a user retrieves a block of $D$ messages with contiguous indices from $K$ replicated messages stored across $N$ non-colluding servers, while hiding the identity of the requested block from each server.
This problem is motivated by storage and streaming systems where files are split into ordered segments.
Unlike multi-message Private Information Retrieval (MPIR), where any $D$-subset may be requested, PCBR restricts the demand family to contiguous blocks. 
This relaxation raises a natural question: Can this structure be exploited to improve retrieval efficiency?
We answer this question for balanced $\{0,1\}$-linear schemes. 
We establish an upper bound on the achievable retrieval rate for all problem parameters, derive a lower bound on the subpacketization level required by any scheme achieving the rate upper bound, and construct a rate-optimal scheme whose subpacketization level matches the lower bound for a broad range of problem parameters. 
Although the optimal PCBR rate coincides with the best-known MPIR rate converse bound, existing MPIR schemes can be suboptimal for PCBR and can require a much larger subpacketization level. 
In contrast, our scheme exploits the contiguous-block structure to achieve the optimal rate with reduced subpacketization.
\end{abstract}

\section{Introduction}
In this paper, we introduce the \emph{Private Contiguous-Block Retrieval (PCBR)} problem. 
In this problem, a database of $K$ messages is replicated on $N$ non-colluding servers, and a user wants to retrieve $D$ out of these $K$ messages whose indices are chosen from a family of candidate size-$D$ interval subsets of ${[1:K]}$, without wrap-around. 
This family is known to the servers, and the user wishes to hide the identity of the desired interval from each server while maximizing the retrieval rate, defined as the ratio of the number of desired message symbols to the total number of retrieved symbols.

This problem is motivated by storage and streaming systems where large files are split into ordered segments, such as video chunks, database pages, or memory blocks, and replicated across servers. 
In such settings, users often request contiguous ranges of segments rather than arbitrary subsets. 
Thus, only segments with consecutive indices are candidate demands, and each server should not be able to identify which contiguous block the user is retrieving.

The PCBR problem belongs to the broad and well-studied class of Private Information Retrieval (PIR) problems~\cite{SJ2017,SJ2016ArbitraryTIFS,SJ2018Multiround,TSC2019,VBU2022}.
Several PIR variants have been studied; see, e.g.,~\cite{VWU2023,UAGJTT2022} and the references therein.
The closest PIR variant to PCBR is multi-message PIR (MPIR).
In MPIR, the query sent to any server must not reveal the user's desired $D$-subset among all $\binom{K}{D}$ possible subsets.
Since the PCBR demand family is contained in this full collection of $D$-subsets, any MPIR scheme also hides the desired subset within the PCBR demand family, and can therefore be applied directly to PCBR~\cite{BU2018,WHS2022,HWS2025,WHS2025,HZTS2025}.

Applying MPIR schemes to PCBR may, however, be suboptimal in terms of the retrieval rate, since MPIR is designed for the full demand space rather than a structured demand family. 
Moreover, many existing MPIR schemes are linear and require messages to be divided into subpackets~\cite{BU2018,HWS2025}. 
A large subpacketization level can increase overhead and implementation complexity. 
By exploiting the structure of the PCBR demand family, it may be possible to reduce subpacketization and obtain simpler, more efficient schemes. 
This motivates the development of retrieval schemes tailored specifically to the PCBR problem.

In this work, we focus on \emph{balanced ${\{0,1\}}$-linear} schemes.
In such schemes, each message is partitioned into the same number of  subpackets; the user queries each server a set of ${\{0,1\}}$-linear combinations of these subpackets, and the server answers the user with the corresponding combinations; and the user sends equal-size queries to all servers and receives equal-size answers in return.
These schemes are practical since they work over any field and yield balanced upload, download, and computation costs across servers. 

We establish a converse bound on the maximum achievable retrieval rate for the PCBR problem and propose a balanced ${\{0,1\}}$-linear PCBR scheme that achieves this bound for all values of $N$, $K$, and $D$.
In addition, we derive a lower bound on the subpacketization level required by balanced ${\{0,1\}}$-linear PCBR schemes and show that our scheme matches this bound for a broad range of values of $N$, $K$, and $D$.

Our results show that the optimal PCBR rate coincides with the best-known MPIR rate converse bound in~\cite{BU2018}.
Nevertheless, when ${D<K/2}$ and ${D\nmid K}$, the best currently known MPIR schemes for this regime, given in~\cite{HWS2025}, do not achieve this bound and therefore are not rate-optimal for PCBR.
In contrast, our proposed PCBR scheme is rate-optimal for all $N$, $K$, and $D$.
When ${D\mid K}$, existing rate-optimal MPIR schemes~\cite{BU2018,HWS2025}, which are balanced ${\{0,1\}}$-linear, can also serve as rate-optimal PCBR schemes. 
However, as shown in~\cite{H2026}, they require a subpacketization level of at least ${N^{K-D+1}/D}$, whereas our PCBR scheme requires the much smaller subpacketization level $N^{K/D}$. 
Finally, when ${D>K/2}$, the rate-optimal MPIR scheme of~\cite{BU2018} can again serve as a rate-optimal PCBR scheme and has the same subpacketization level $N^2$ as our scheme. 
However, it requires a sufficiently large field size, whereas our scheme works over any field. 

\section{Problem Setup}
For any integers ${i,j}$ with ${0\leq i\leq j}$, we denote the set ${\{i,i+1,\dots,j\}}$ by ${[i:j]}$. 
We denote random variables by bold-face symbols and their realizations by regular symbols.
We fix an arbitrary prime power $q$ throughout, denote the finite field of order $q$ by $\mathbb{F}_q$, and denote the $L$-dimensional vector space over $\mathbb{F}_q$ by $\mathbb{F}_q^L$ for any integer $L\geq 1$.
All entropy and mutual information quantities are measured in $q$-ary units.

A dataset consisting of $K$ messages ${\mathrm{X}_1,\dots,\mathrm{X}_K}$, is replicated across ${N}$ non-colluding servers. 
Each message $\mathrm{X}_i\in \mathbb{F}_{q}^{L}$, for ${i \in [1:K]}$, consists of $L$ symbols from ${\mathbb{F}_q}$.

To simplify the notation, for any ${\mathrm{U} \subseteq [1:K]}$, we define \[{\mathrm{X}_{\mathrm{U}} \coloneqq \{\mathrm{X}_i: i\in \mathrm{U}\}}.\]

A user wishes to retrieve $D$ messages, for some ${D\in[2:K-1]}$, indexed by $\mathrm{W}\in \{\mathrm{W}_1,\dots,\mathrm{W}_E\}$, where ${E\coloneqq K-D+1}$ and, for each ${j\in [1:E]}$, 
\[{\mathrm{W}_j \coloneqq [j:j+D-1]}.\]
Equivalently, $\mathrm{W}_1,\dots,\mathrm{W}_E$ are the $E$ distinct size-$D$ interval subset of ${[1:K]}$, without wrap-around.  

We refer to $\mathrm{X}_{\mathrm{W}}$ as the \emph{demand messages},  ${\mathrm{X}_{[1:K]\setminus \mathrm{W}}}$ as the \emph{interference messages}, $\mathrm{W}$ as the \emph{demand index set}, and ${\mathrm{W}_1,\dots,\mathrm{W}_E}$ as the \emph{candidate demand index sets}.

We assume that the random variables ${\mathbf{X}_1,\dots,\mathbf{X}_K}$ are independent and uniformly distributed over ${\mathbb{F}_{q}^{L}}$. 
We also assume that the random variable $\mathbf{W}$ is distributed arbitrarily over $\{\mathrm{W}_1,\dots,\mathrm{W}_E\}$ such that every $\mathrm{W}_j$ for ${j\in [1:E]}$ has a nonzero probability. 
In addition, we assume that 
${\mathbf{X}_{[1:K]}}$ and $\mathbf{W}$ are independent random variables.

The user constructs $N$ queries ${\mathrm{Q}^{[\mathrm{W}]}_{1},\ldots,\mathrm{Q}^{[\mathrm{W}]}_{N}}$ and sends $\mathrm{Q}^{[\mathrm{W}]}_{n}$ to server $n$ for each ${n\in [1:N]}$; 
each query $\mathrm{Q}^{[\mathrm{W}]}_{n}$ is a (possibly stochastic) function of $\mathrm{W}$ and is independent of $\mathrm{X}_{[1:K]}$. 
That is, 
\begin{equation*}
I(\mathbf{Q}_{[1:N]}^{[\mathrm{W}]};\mathbf{X}_{[1:K]})=0,
\end{equation*} 
where $\mathbf{Q}_{[1:N]}^{[\mathrm{W}]}\coloneqq \{\mathbf{Q}^{[\mathrm{W}]}_{1},\dots,\mathbf{Q}^{[\mathrm{W}]}_{N}\}$. 

Upon receiving the query ${\mathrm{Q}_{n}^{[\mathrm{W}]}}$, each server $n$ generates an answer ${\mathrm{A}_{n}^{[\mathrm{W}]}}$ and sends it back to the user;  
each answer $\mathrm{A}^{[\mathrm{W}]}_{n}$ is a deterministic function of $\mathrm{Q}^{[\mathrm{W}]}_{n}$ and $\mathrm{X}_{[1:K]}$, i.e., 
\begin{equation*} 
H(\mathbf{A}_{n}^{[\mathrm{W}]} | \mathbf{Q}_{n}^{[\mathrm{W}]}, \mathbf{X}_{[1:K]})=0, \quad \forall n \in [1:N].
\end{equation*}

After receiving the answers from all servers, the user must be able to recover $\mathrm{X}_{\mathrm{W}}$, i.e.,
\begin{equation} 
\label{eq:correctness}
H( \mathbf{X}_{\mathrm{W}} | \mathbf{Q}_{[1:N]}^{[\mathrm{W}]}, \mathbf{A}_{[1:N]}^{[\mathrm{W}]})=0,
\end{equation} where $\mathbf{A}_{[1:N]}^{[\mathrm{W}]}\coloneqq \{\mathbf{A}^{[\mathrm{W}]}_{1},\dots,\mathbf{A}^{[\mathrm{W}]}_{N}\}$. 
This requirement is referred to as the \emph{correctness condition}. 

For each server ${n\in [1:N]}$, its observations $\mathrm{Q}^{[\mathrm{W}]}_n$, $\mathrm{A}^{[\mathrm{W}]}_n$, and $\mathrm{X}_{[1:K]}$, must not reveal any information about $\mathrm{W}$. 
That is,
\begin{equation}
\label{eq:privacy}	I(\mathbf{W};\mathbf{Q}_{n}^{[\mathrm{W}]},\mathbf{A}_{n}^{[\mathrm{W}]},\mathbf{X}_{[1:K]})=0, \quad \forall n \in [1:N].
\end{equation}
We refer to this requirement as the \emph{privacy condition}.

The problem is to design a scheme that satisfies the correctness condition in~\eqref{eq:correctness} and the privacy condition in~\eqref{eq:privacy}. 
We refer to this problem as \emph{Private Contiguous-Block Retrieval (PCBR)}. 
This problem is a special case of the Private Structured-Subset Retrieval (PSSR) problem introduced in our parallel work~\cite{IH2026PSSRarXiv} with the candidate demand index sets given by those specified earlier. 

In this work, we focus on a class of PCBR schemes that we call \emph{balanced $\{0,1\}$-linear PCBR schemes}. 
In such schemes, each message is partitioned into $L$ subpackets, each consisting of a single message symbol. The user queries each server for a collection of linear combinations of message subpackets with coefficients in $\{0,1\}$, and
each server answers the user with the corresponding linear combinations.
Moreover, the queries sent to all servers have the same total length, and the
answers returned by all servers have the same total length.

To evaluate the performance of a scheme, we consider two metrics: 
the \emph{retrieval rate} and the \emph{subpacketization level}. 
The \emph{retrieval rate} is defined as the ratio of the amount of information the user requires, namely ${H(\mathbf{X}_{\mathbf{W}})}$, to the total amount of information retrieved from all servers, namely ${\sum_{n=1}^N H (\mathbf{A}_{n}^{[\mathbf{W}]} | \mathbf{Q}_{n}^{[\mathbf{W}]})}$, and the \emph{subpacketization level} is defined as the number of subpackets into which each message is partitioned by the scheme.

Our goal in this work is twofold: 
\begin{itemize}
\item[(i)] to establish converse bounds, as functions of the number of servers $N$, the total number of messages $K$, and the number of demand messages $D$, on the maximum retrieval rate achievable by balanced $\{0,1\}$-linear PCBR schemes and on the minimum subpacketization level required to achieve this rate; and 
\item[(ii)] to design schemes that achieve these bounds, or closely approach them.
\end{itemize}

\section{Main results}
In this section, we present our main converse and achievability results for the PCBR problem. 

To simplify the notation, we define 
\[f \coloneqq \left\lfloor \frac{K}{D}\right\rfloor, \quad g \coloneqq \left\lceil \frac{K}{D}\right\rceil.\]

\begin{theorem}
\label{thm:PRSR_cons_dem}
For $N$ servers, a total of $K$ messages, and $D$ demand messages, the maximum retrieval rate achievable by balanced $\{0,1\}$-linear PCBR schemes is 
\begin{equation}
\begin{aligned}
    \label{eq:PSSR_cap_cons_dem}
R &\coloneqq \frac{DN^{f} }{DN (N^{f}-1)/(N-1)+K-Df}.
\end{aligned}
\end{equation}
\end{theorem}

The converse follows  from~\cite[Theorem~1]{IH2026PSSRarXiv}, which upper bounds the retrieval rate of any scheme for the PSSR problem, and in particular of any balanced ${\{0,1\}}$-linear scheme, and hence applies to the PCBR setting as a special case. 

To establish achievability, we propose a balanced ${\{0,1\}}$-linear PCBR scheme whose retrieval rate matches the converse bound.  
Although we present it here as a standalone scheme, it is instantiated from the optimization-based framework that we developed in~\cite{IH2026PSSRarXiv} for the general PSSR setting, and specialized to the interval-demand family considered here by exploiting the symmetry among the candidate demand index sets.   

\begin{theorem}\label{thm:PRSR_cons_L}
For $N$ servers, a total of $K$ messages, and $D$ demand messages, the minimum subpacketization level required by any balanced $\{0,1\}$-linear PCBR scheme achieving the rate $R$ in~\eqref{eq:PSSR_cap_cons_dem} is lower bounded by
\begin{equation}
\begin{aligned}\label{eq:PCBR_lb_L}
L_{*}\coloneqq \frac{N^{g}}{\gcd(N^{g}, D(N^{g}-1)/(N-1) + K-Dg)},
\end{aligned}
\end{equation} 
and is upper bounded by  
\begin{equation}\label{eq:PCBR_ub_L}
L^{*}\coloneqq N^{g}.
\end{equation}
The bounds coincide when $N$ and $K-D(g-1)$ are coprime. 
\end{theorem}

The lower bound is an immediate consequence of the integrality of the number of linear combinations the user retrieves from each server in any balanced $\{0,1\}$-linear scheme, and 
the upper bound is achieved by the proposed PCBR scheme. 

\begin{remark}
\normalfont
Any MPIR scheme is a valid PCBR scheme, since in the MPIR setting every $D$-subset of ${[1:K]}$ is a candidate demand index set, whereas the PCBR setting restricts the candidate demand index sets to size-$D$ interval subsets of ${[1:K]}$. 
Moreover, for the PCBR problem the maximum achievable retrieval rate, namely the rate $R$ in~\eqref{eq:PSSR_cap_cons_dem}, matches the best-known upper bound for the MPIR problem, originally established in~\cite{BU2018} and later recovered in~\cite{IH2026PSSRarXiv} as a special case. 
This, however, does not imply that the best-known MPIR schemes are always rate-optimal for PCBR. 
Notably, in the regime ${D < K/2}$ and ${D \nmid K}$, the highest known achievable MPIR rate is due to the scheme of~\cite{HWS2025}, which is also a balanced ${\{0,1\}}$-linear scheme; 
however, the scheme of~\cite{HWS2025} achieves a rate strictly smaller than
$R$, and hence it is not rate-optimal for PCBR, in contrast to our scheme which achieves optimal retrieval rate in this regime.   
\end{remark}

\begin{remark}
\normalfont
For the regime ${D \mid K}$, the MPIR scheme of~\cite{BU2018}, which is also balanced ${\{0,1\}}$-linear, is rate-optimal for PCBR; 
however, it has been shown in~\cite{H2026} that this scheme requires a subpacketization level of at least $N^{K-D+1}/D$ , which can be significantly larger than the subpacketization level $N^{K/D}$ required by our rate-optimal scheme in this regime. 
For the regime ${D > K/2}$,~\cite{BU2018} presents a different MPIR scheme that is also rate-optimal for PCBR and requires the optimal subpacketization level $N^{2}$; 
however, this scheme is not ${\{0,1\}}$-linear and relies on maximum distance separable (MDS) codes, and thus applies only when the field size is sufficiently large. 
In contrast, our scheme is
rate-optimal in this regime while remaining balanced ${\{0,1\}}$-linear and
requiring the optimal subpacketization level $N^{2}$.
\end{remark}

\section{Converse Proofs}
In this section, we present the converse proofs for Theorem~\ref{thm:PRSR_cons_dem}, which establishes the upper bound $R$ in~\eqref{eq:PSSR_cap_cons_dem} on the maximum achievable retrieval rate, and for Theorem~\ref{thm:PRSR_cons_L}, which establishes the lower bound $L_{*}$ in~\eqref{eq:PCBR_lb_L} on the minimum subpacketization level.

\subsection{Upper Bounding the Achievable Rate}
Since PCBR is a special case of PSSR with ${E = K-D+1}$ candidate demand index sets ${\mathrm{W}_1,\dots,\mathrm{W}_E}$, where 
${\mathrm{W}_j = [j,j+D-1]}$
for all ${j\in [1:E]}$, it follows from~\cite[Theorem~1]{IH2026PSSRarXiv} that, for every permutation $\pi: [1:E]\rightarrow [1:E]$, the maximum achievable retrieval rate is upper bounded by
\begin{align}
\label{eq:PSR ub}
D \left(
 \sum_{j=1}^{E} \frac{1}{N^{j-1}} \left| {\mathrm{W}_{\pi(j)}} \setminus \textstyle\bigcup_{k=0}^{j-1} {\mathrm{W}_{\pi(k)}} \right| \right)^{-1},
\end{align} where ${\pi(0)\coloneqq 0}$ and ${\mathrm{W}_{0}\coloneqq \emptyset}$. 

Take $\pi$ such that ${\pi(j) = (j-1)D+1}$ for all ${j\in [1:f]}$, ${\pi(f+1) = K-D+1}$, and $\pi(j)$ is arbitrary for ${j\in [f+2:E]}$, where ${f = \lfloor K/D \rfloor}$. 
Then, ${\mathrm{W}_{\pi(1)},\dots,\mathrm{W}_{\pi(f)}}$ are pairwise disjoint $D$-subsets of ${[1:K]}$, and $\mathrm{W}_{\pi(f+1)}$ contains all remaining ${K-Df}$ indices in ${[1:K]}$ not covered by ${\mathrm{W}_{\pi(1)},\dots,\mathrm{W}_{\pi(f)}}$.  
Substituting into~\eqref{eq:PSR ub} yields
\begin{equation*}
D\left(\sum_{j=1}^{f} \frac{D}{N^{j-1}}+\frac{K-Df}{N^{f}} \right)^{-1},
\end{equation*} 
which further simplifies to $R$ in~\eqref{eq:PSSR_cap_cons_dem}.  

\subsection{Lower Bounding the Subpacketization Level}
For any balanced ${\{0,1\}}$-linear PCBR scheme achieving the rate $R$ in~\eqref{eq:PSSR_cap_cons_dem}, the quantity $DL/(NR)$, which represents the number of linear combinations the user retrieves from each server, must be an integer. 

When ${D \mid K}$, we have ${f=g}$, so
\[
\frac{DL}{NR} = \frac{L(D(N^g-1)/(N-1))}{N^g},
\]
and hence the smallest positive integer $L$ is 
\[
\frac{N^g}{\gcd(N^g,D(N^g-1)/(N-1))}, 
\]
which agrees with $L_{*}$ in~\eqref{eq:PCBR_lb_L}.

When ${D \nmid K}$, we have ${f=g-1}$, and hence
\begin{align*}
\frac{DL}{NR} & = \frac{L(DN(N^{g-1}-1)/(N-1)+D+K-Dg)}{N^{g}}\\
& = \frac{L(D(N^{g}-1)/(N-1)+K-Dg)}{N^{g}},
\end{align*}
so the smallest positive integer $L$ is
\[
\frac{N^g}{\gcd(N^g,D(N^{g}-1)/(N-1)+K-Dg)} 
\]
which again matches $L_{*}$ in~\eqref{eq:PCBR_lb_L}.

\section{Achievability Proofs}
In this section, we present a balanced ${\{0,1\}}$-linear PCBR scheme that achieves the rate $R$ in~\eqref{eq:PSSR_cap_cons_dem} with subpacketization level $L^{*}$ in~\eqref{eq:PCBR_ub_L}. 
This establishes the achievability parts of Theorems~\ref{thm:PRSR_cons_dem} and~\ref{thm:PRSR_cons_L}.

Each message is partitioned into ${L^{*}=N^g}$ subpackets, where ${g = \lceil K/D\rceil}$, and each subpacket is a randomly chosen $\mathbb{F}_q$-symbol from the message.
For each server ${n\in[1:N]}$, the user sends a query $\mathrm{Q}^{[\mathrm{W}]}_n$, which specifies a collection of symbols (with the same number of symbols for all $n$). 
For every fixed set of messages, each server is queried for the same number of symbols, i.e., $\{0,1\}$-linear combinations involving subpackets of those messages.
Each message contributes at most one subpacket to each symbol, and no subpacket appears more than once among the symbols retrieved from any given server.
Server $n$ returns the answer $\mathrm{A}^{[\mathrm{W}]}_n$, which consists of the corresponding symbols obtained by forming the specified ${\{0,1\}}$-linear combinations over $\mathbb{F}_q$.

For each non-empty subset ${\mathrm{U}\subseteq[1:K]}$, let $T_{\mathrm{U}}$ denote the number of symbols that the user retrieves from each server and whose support is exactly $\mathrm{U}$, i.e., symbols that involve precisely the messages indexed by $\mathrm{U}$.

We present the proposed scheme separately for the two regimes ${D\leq K/2}$ and ${D>K/2}$.

\subsection{Achievable Scheme for ${D\leq K/2}$}

We define
\begin{equation*}
M \coloneqq K-D(g-1) =
\begin{cases}
D, & \text{if } D \mid K,\\
K-Df, & \text{otherwise,}
\end{cases}
\end{equation*} 
where ${f = \lfloor K/D\rfloor}$. 

We also define a partition ${\mathrm{S}_1\sqcup \mathrm{S}_2}$ of ${[1:K]}$ with an alternating block structure.
Specifically, $\mathrm{S}_1$ is partitioned into $g$ blocks $\mathrm{S}_{1,1},\dots,\mathrm{S}_{1,g}$, each of size $M$, and $\mathrm{S}_2$ is partitioned into $f$ blocks $\mathrm{S}_{2,1},\dots,\mathrm{S}_{2,f}$, each of size $D-M$, i.e.,
\[
\mathrm{S}_1 \coloneqq \bigsqcup_{l=1}^{g} \mathrm{S}_{1,l},
\qquad
\mathrm{S}_2 \coloneqq \bigsqcup_{l=1}^{f} \mathrm{S}_{2,l},
\]
with ${|\mathrm{S}_{1,l}|=M}$ for all ${l\in[1:g]}$ and ${|\mathrm{S}_{2,l}|=D-M}$ for all ${l\in[1:f]}$. 
Note that ${|\mathrm{S}_1|=Mg}$ and ${|\mathrm{S}_2|=(D-M)f}$, so 
\[{|\mathrm{S}_1|+|\mathrm{S}_2| = Mg+(D-M)f = M(g-f)+ Df= K},\] 
since ${f=g}$ when $D\mid K$, and ${f=g-1}$ when ${D\nmid K}$. 

These blocks appear along ${[1:K]}$ in an alternating order: 
$\mathrm{S}_{1,1}$ consists of the first $M$ indices ${[1:M]}$, $\mathrm{S}_{2,1}$ consists of the next ${D-M}$ indices ${[M+1:D]}$, and so on; 
in particular, the last block of $\mathrm{S}_1$ is ${\mathrm{S}_{1,g}=[K-M+1:K]}$.

When ${D\mid K}$, we have ${M=D}$ and hence ${\mathrm{S}_2=\emptyset}$, whereas when ${D\nmid K}$, we have ${M<D}$ and hence ${\mathrm{S}_2\neq \emptyset}$.

With these definitions in place, we now describe the proposed scheme.

From each server, the user retrieves two types of symbols:
\emph{singleton symbols}, each containing a single message subpacket, and
\emph{$k$-sum symbols}, for ${k\geq 2}$, each formed by summing $k$ subpackets from $k$ distinct messages.
The numbers of retrieved symbols of each type are specified below.

\begin{itemize}
    \item \emph{Singleton symbols.}
    For each ${i\in[1\!:\!K]}$, we have 
    \[
        T_{\{i\}}=
        \begin{cases}
        1, & i\in \mathrm{S}_1,\\
        N, & i\in \mathrm{S}_2.
        \end{cases}
    \]
    \item \emph{$k$-sum symbols from $\mathrm{S}_1$ ($k\in[2:g]$).}
    For any ${\mathrm{U}\subseteq \mathrm{S}_1}$ with ${|\mathrm{U}|=k}$, we have \[T_{\mathrm{U}}=(N-1)^{k-1}\] 
    if and only if $\mathrm{U}$ contains exactly one element from each of $k$ distinct blocks of $\mathrm{S}_1$, say ${\mathrm{S}_{1,l_1},\dots,\mathrm{S}_{1,l_k}}$, and all these elements are congruent modulo $D$. 
    Equivalently, there exist distinct indices ${l_1,\dots,l_k\in[1:g]}$ and elements ${i\in \mathrm{S}_{1,l_1}}$ and ${j_t\in \mathrm{S}_{1,l_t}}$ for ${t\in[2:k]}$ such that ${j_t\equiv i \pmod D}$ for all ${t\in[2:k]}$ and ${\mathrm{U}=\{i,j_2,\dots,j_k\}}$. 
    Otherwise, ${T_{\mathrm{U}}=0}$.
    \item \emph{$k$-sum symbols from $\mathrm{S}_2$ ($k\in[2:f]$).}
    For any ${\mathrm{U}\subseteq \mathrm{S}_2}$ with ${|\mathrm{U}|=k}$, we have  \[{T_{\mathrm{U}}=N(N-1)^{k-1}}\] if and only if $\mathrm{U}$ contains exactly one element from each of $k$ distinct blocks of $\mathrm{S}_2$, say ${\mathrm{S}_{2,l_1},\dots,\mathrm{S}_{2,l_k}}$, and all these elements are congruent modulo $D$. 
    Equivalently, there exist distinct ${l_1,\dots,l_k\in[1:f]}$ and elements ${i\in \mathrm{S}_{2,l_1}}$ and ${j_t\in \mathrm{S}_{2,l_t}}$ for ${t\in[2:k]}$ such that ${j_t\equiv i \pmod D}$ for all ${t\in[2:k]}$ and ${\mathrm{U}=\{i,j_2,\dots,j_k\}}$. Otherwise, ${T_{\mathrm{U}}=0}$.
    \item \emph{Mixed $k$-sum symbols.}
    For any ${\mathrm{U}}$ that contains elements from both $\mathrm{S}_1$ and $\mathrm{S}_2$, we have ${T_{\mathrm{U}}=0}$.
\end{itemize}

We next describe the subpacket indexing procedure for each retrieved symbol.

Subpacket indices are assigned sequentially in increasing sum size.
The assignment starts with singleton symbols (${k=1}$): 
for each message $\mathrm{X}_i$, the $T_{\{i\}}$ singleton symbols of $\mathrm{X}_i$ at each server are given distinct subpacket indices that have not been used previously at any server. 
The procedure then continues with $k$-sum symbols for ${k\geq 2}$.

Each $k$-sum symbol involves subpackets from at most one demand message. 
Indeed, the message indices appearing in any $k$-sum symbol satisfy ${j_t \equiv i \pmod D}$, whereas the demand message indices form a size-$D$ consecutive set. 
Thus, for a fixed $i$, at most one demand index can satisfy ${j \equiv i \pmod D}$. 
Accordingly, $k$-sum symbols are grouped into two categories: 
those involving only interference messages, and 
those containing exactly one demand message. 
The indexing rules for these two categories are as follows:
\begin{itemize}
\item For a $k$-sum symbol involving only interference messages, each participating message is assigned an index that has not been used previously at any server.
\item For a $k$-sum symbol at a given server with support $\mathrm{U}$ that includes a demand message $\mathrm{X}_i$, the message $\mathrm{X}_i$ is assigned an index that has not been used previously at any server, while each interference message in ${\mathrm{U}\setminus\{i\}}$ is assigned the \emph{same} index as in a ${(k-1)}$-sum symbol with support ${\mathrm{U}\setminus\{i\}}$ at another server, chosen so that this reused index has not been used previously at the given server.
\end{itemize}

This concludes our description of the subpacket indexing procedure and, in turn, of the proposed scheme.

\subsection{Achievable Scheme for ${D>K/2}$}
When ${D>K/2}$, any size-$D$ interval subset of $[1:K]$ contains a common set of ${2D-K}$ message indices, namely ${[K-D+1:D]}$.

In this regime, we have ${g=2}$, so the scheme uses subpacketization level ${L^{*} = N^{g}=N^{2}}$ and proceeds in two phases. 

The first phase involves retrieving the messages included in every candidate demand set. 
Specifically, the user directly retrieves $N$ distinct subpackets of each of these messages from each server.

In the second phase, the user privately retrieves the remaining ${\widehat{D}\coloneqq K-D}$ demand messages from the remaining ${\widehat{K}\coloneqq 2K-2D}$ messages. 
To do so, since ${\widehat{D}=\widehat{K}/2}$, the user applies the scheme for the regime ${D\leq K/2}$ to this reduced instance with parameters ${(\widehat{K},\widehat{D})}$.

\subsection{Proof of Achievable Rate}
We now compute the rate achieved by the scheme in each regime and show that it is equal to $R$ in~\eqref{eq:PSSR_cap_cons_dem}.

We begin with the scheme for the regime ${D\leq K/2}$.
Recall that, in this regime, for each non-empty ${\mathrm{U}\subseteq[1:K]}$, the user retrieves from each server $T_{\mathrm{U}}$ symbols whose support is $\mathrm{U}$.

\emph{Symbols supported on $\mathrm{S}_1$.}
Since $\mathrm{S}_1$ is partitioned into $g$ blocks ${\mathrm{S}_{1,1},\dots,\mathrm{S}_{1,g}}$ of size $M$, we have ${|\mathrm{S}_1| = Mg}$.
By construction, for every ${i\in \mathrm{S}_1}$ we have ${T_{\{i\}}=1}$, so the user retrieves ${Mg}$ singleton symbols from each server whose support is contained in $\mathrm{S}_1$.

Next, consider the $k$-sum symbols for each ${k\in[2:g]}$ whose support is contained in $\mathrm{S}_1$.
Fix any $k$ distinct blocks ${\mathrm{S}_{1,l_1},\dots,\mathrm{S}_{1,l_k}}$. 
For each position ${m\in[1:M]}$ within a block, there is a unique index in each chosen block that occupies position $m$. 
Selecting these $k$ indices---one from each of the $k$ blocks, all at the same position $m$---forms a $k$-subset ${\mathrm{U}\subseteq \mathrm{S}_1}$ that satisfies the condition in the scheme description, and every subset ${\mathrm{U}\subseteq \mathrm{S}_1}$ with ${T_{\mathrm{U}}\neq 0}$ is generated in this manner. 
Thus, for a fixed choice of $k$ blocks there are exactly $M$ such supports $\mathrm{U}$, and over all choices of $k$ blocks the number of supports ${\mathrm{U}\subseteq \mathrm{S}_1}$ with ${T_{\mathrm{U}}\neq 0}$ is ${M\binom{g}{k}}$. 
Moreover, for each such $\mathrm{U}$, the user retrieves \[{T_{\mathrm{U}}=(N-1)^{k-1}}\] $k$-sum symbols supported on $\mathrm{U}$ from each server. 
Therefore, the total number of $k$-sum symbols with support contained in $\mathrm{S}_1$ that the user retrieves from each server is \[{M\binom{g}{k}(N-1)^{k-1}}.\]

Summing over $k$, the total number of symbols that the user retrieves from each server and whose support is contained in $\mathrm{S}_1$ is given by
\begin{align}
& Mg+M\sum_{k=2}^{g} \binom{g}{k}(N-1)^{k-1}\nonumber \\
& \quad = M\sum_{k=1}^{g}\binom{g}{k}(N-1)^{k-1}\nonumber\\
& \quad = M(N^{g}-1)/(N-1), \label{eq:nbr_of_lin_comb_set1}
\end{align}
where~\eqref{eq:nbr_of_lin_comb_set1} follows from the binomial theorem.

\emph{Symbols supported on $\mathrm{S}_2$.}
Recall that $\mathrm{S}_2$ is partitioned into $f$ blocks ${\mathrm{S}_{2,1},\dots,\mathrm{S}_{2,f}}$, each of size ${D-M}$, so ${|\mathrm{S}_2|=(D-M)f}$.
For every ${i\in \mathrm{S}_2}$ we have ${T_{\{i\}}=N}$, hence the number of singleton symbols from $\mathrm{S}_2$ that the user retrieves from each server equals ${|\mathrm{S}_2|N=(D-M)fN}$.

Next consider the $k$-sum symbols for each ${k\in[2:f]}$ whose support is contained in $\mathrm{S}_2$.
Fix any $k$ distinct blocks ${\mathrm{S}_{2,l_1},\dots,\mathrm{S}_{2,l_k}}$. 
For each position ${m\in[1:D-M]}$ within a block, there is a unique index in each chosen block that occupies position $m$. 
Taking these $k$ indices (one from each of the $k$ blocks, all at the same position $m$) produces a $k$-subset ${\mathrm{U}\subseteq \mathrm{S}_2}$ with ${T_{\mathrm{U}}\neq 0}$, and every ${\mathrm{U}\subseteq \mathrm{S}_2}$ with ${T_{\mathrm{U}}\neq 0}$ is obtained in this manner. 
Thus, for a fixed choice of $k$ blocks there are exactly ${D-M}$ such supports $\mathrm{U}$, and over all choices of $k$ blocks the number of supports ${\mathrm{U}\subseteq \mathrm{S}_2}$ with ${T_{\mathrm{U}}\neq 0}$ is ${(D-M)\binom{f}{k}}$. 
Moreover, for each such $\mathrm{U}$, the user retrieves \[{T_{\mathrm{U}}=N(N-1)^{k-1}}\] $k$-sum symbols supported on $\mathrm{U}$ from each server. 
Thus, the total number of $k$-sum symbols with support contained in $\mathrm{S}_2$ that the user retrieves from each server is \[{(D-M)\binom{f}{k}\,N(N-1)^{k-1}}.\]

Summing over $k$, the total number of symbols that the user retrieves from each server and whose support is contained in $\mathrm{S}_2$ is given by
\begin{align}
& (D-M)fN+\sum_{k=2}^{f}(D-M)\binom{f}{k}N(N-1)^{k-1} \nonumber \\
& \quad = (D-M)N \sum_{k=1}^{f}\binom{f}{k}(N-1)^{k-1} \nonumber\\
& \quad = (D-M)N(N^{f}-1)/(N-1), \label{eq:nbr_of_lin_comb_set2}
\end{align}
where~\eqref{eq:nbr_of_lin_comb_set2} again follows from the binomial theorem.

Since each message is partitioned into ${L^{*} = N^{g}}$ subpackets and the demand consists of $D$ messages, there are ${DN^{g}}$ demand subpackets, each an $\mathbb{F}_q$-symbol. 
Moreover, each server returns a collection of symbols to the user, each an element of $\mathbb{F}_q$, and this collection has the same cardinality for all servers. 
Thus, the total number of symbols the user retrieves equals $N$ times the number of symbols retrieved from a single server, and the rate is the ratio of ${DN^{g}}$ to this total number.

When ${D\mid K}$, we have ${M=D}$ and ${f=g=K/D}$, so ${\mathrm{S}_2=\emptyset}$ and the user retrieves from each server only symbols supported on $\mathrm{S}_1$. 
Using~\eqref{eq:nbr_of_lin_comb_set1}, the number of such symbols is ${D(N^{g}-1)/(N-1)}$, and hence the rate equals
\begin{align*}
\frac{DN^{g}}{ND(N^{g}-1)/(N-1)}
= \frac{N^{g}(N-1)}{N(N^{g}-1)}. \end{align*}
This matches the rate $R$ in~\eqref{eq:PSSR_cap_cons_dem} for the case ${D\mid K}$.

When ${D\nmid K}$, we have ${f=g-1}$ and ${M=K-Df}$. 
In this case, ${\mathrm{S}_2\neq \emptyset}$, and the user retrieves from each server symbols supported on $\mathrm{S}_1$ and on $\mathrm{S}_2$. 
Using~\eqref{eq:nbr_of_lin_comb_set1} and~\eqref{eq:nbr_of_lin_comb_set2}, the numbers of such symbols are
${M(N^{g}-1)/(N-1)}$ and ${(D-M)N(N^{f}-1)/(N-1)}$, respectively. 
Summing these, the total number of symbols the user retrieves from each server equals $DN(N^{f}-1)/(N-1)+K-Df$, and hence the rate is 
\begin{align}
& \frac{DN^{g}}{N\left(DN(N^{f}-1)/(N-1)+K-Df\right)} \nonumber \\
& \quad = \frac{DN^f}{DN(N^f-1)/(N-1)+K-Df},\label{eq:rate_cons_demands_D_not|K}
\end{align} 
which coincides with the rate $R$ in~\eqref{eq:PSSR_cap_cons_dem} for the case ${D\nmid K}$. 

We next consider the scheme for the regime ${D>K/2}$. 
Recall that, in this regime, ${g=2}$ and hence the subpacketization level is ${L^{*} = N^{2}}$, and that the scheme operates in two phases.

The first phase directly retrieves ${(2D-K)N^{2}}$ demand subpackets, each an $\mathbb{F}_q$-symbol. 
The second phase applies the scheme for the regime ${D\leq K/2}$ to a reduced instance with parameters $({\widehat{K}, \widehat{D}) = (2K-2D,K-D)}$; 
since ${\widehat{D}\mid \widehat{K}}$ and ${\widehat{g}= \widehat{K}/\widehat{D}=2}$, this phase retrieves a total of
\[{N\widehat{D} \left(\frac{N^{\widehat{g}}-1}{N-1}\right)=N(K-D)(N+1)}\] symbols, each again an element of $\mathbb{F}_q$. 
Thus, the rate is
\begin{align*}
& \frac{DN^2}{N(K-D)(N+1)+(2D-K)N^2} = \frac{DN}{DN+K-D},
\end{align*} which equals~\eqref{eq:rate_cons_demands_D_not|K} since ${f = g-1 = 1}$ in this regime, and hence matches the rate $R$ in~\eqref{eq:PSSR_cap_cons_dem}. 

\subsection{Proof of Correctness}
We next establish correctness for the scheme in the regime ${D\leq K/2}$. 
This also yields correctness for the scheme in the regime ${D>K/2}$: 
the first phase recovers the common demand messages explicitly, and the second phase completes recovery by applying the same scheme as in the regime ${D\leq K/2}$.

Fix a demand message $\mathrm{X}_i$ and fix a server.
Consider any $k$-sum symbol at this server with support $\mathrm{U}$ satisfying ${i\in \mathrm{U}}$.
By construction, $\mathrm{U}$ contains no other demand message index, and hence the remaining ${k-1}$ messages involved are all interference messages.
The subpacket indexing is chosen so that, for each such $k$-sum symbol at the fixed server, the ${k-1}$ interference subpackets coincide with those in some ${(k-1)}$-sum symbol at another server supported on ${\mathrm{U}\setminus\{i\}}$.
Subtracting that ${(k-1)}$-sum symbol cancels the interference and recovers one subpacket of $\mathrm{X}_i$.
Thus, it remains only to verify that, for every $k$-sum support $\mathrm{U}$ such that ${i\in \mathrm{U}}$, the collection of ${(k-1)}$-sum symbols supported on ${\mathrm{U}\setminus\{i\}}$ across the other servers is large enough to support this cancellation for all $k$-sum symbols supported on $\mathrm{U}$ at the fixed server.

First, consider a demand message $\mathrm{X}_i$ with ${i \in \mathrm{S}_1}$.
The user retrieves from each server one singleton symbol of $\mathrm{X}_i$.
Next, consider the $k$-sum symbols (${k \in [2:g]}$) that include messages whose indices belong to $\mathrm{S}_1$. 
Since $\mathrm{S}_1$ consists of $g$ blocks and a fixed message is combined with at most one message from each block, the message index $i$ appears in $\binom{g-1}{k-1}$ distinct $k$-sum supports. 
Moreover, for each such support, the user retrieves $(N-1)^{k-1}$ $k$-sum symbols from each server. 
Thus, the number of subpackets of $\mathrm{X}_i$ that appear in the symbols retrieved from each server equals
\begin{align}
& 1 + \sum_{k=2}^{g} \binom{g-1}{k-1}(N-1)^{k-1} = N^{g-1}. \nonumber
\end{align}

Fix a server and a $k$-sum support $\mathrm{U}$ with ${i\in \mathrm{U}}$. 
The user retrieves $(N-1)^{k-2}$ ${(k-1)}$-sum symbols supported on ${\mathrm{U}\setminus\{i\}}$ from each of the other servers.
Thus, across the other ${N-1}$ servers there are ${(N-1)\cdot (N-1)^{k-2}=(N-1)^{k-1}}$ available ${(k-1)}$-sum symbols supported on ${\mathrm{U}\setminus\{i\}}$, which matches the number of $k$-sum symbols ${(N-1)^{k-1}}$ supported on $\mathrm{U}$ that the user retrieves from the fixed server.
Therefore, every $k$-sum symbol involving $\mathrm{X}_i$ yields one recoverable subpacket of $\mathrm{X}_i$, and all $N^{g-1}$ subpackets of $\mathrm{X}_i$ appearing in one server are recoverable.
Consequently, across all $N$ servers, the total number of recoverable subpackets of
$\mathrm{X}_i$ is ${N\cdot N^{g-1}=N^{g}}$, which matches the subpacketization level ${L^{*} = N^{g}}$ used by the scheme. 
This completes the proof of recoverability for each demand message $\mathrm{X}_i$ with ${i\in \mathrm{S}_1}$.

Now, consider a demand message $\mathrm{X}_i$ with ${i \in \mathrm{S}_2}$. 
Recall that $\mathrm{S}_2 \neq \emptyset$ only when ${D \nmid K}$, in which case ${f=g-1}$. 
From each server the user retrieves $N$ singleton symbols of $\mathrm{X}_i$.
Next, consider the $k$-sum symbols (${k \in [2:f]}$) that involve messages whose indices belong to $\mathrm{S}_2$. 
Since $\mathrm{S}_2$ consists of $f$ blocks and $\mathrm{X}_i$ is combined with at most one message from each block, the message index $i$ appears in $\binom{f-1}{k-1}$ distinct $k$-sum supports. 
Moreover, for each such support, the user retrieves ${N(N-1)^{k-1}}$ $k$-sum symbols from each server. 
Thus, the number of subpackets of $\mathrm{X}_i$ that appear in the symbols retrieved from each server equals
\begin{align}
& N + \sum_{k=2}^{f} \binom{f-1}{k-1} N(N-1)^{k-1} = N^{f}. \nonumber 
\end{align}

Fix a server and a $k$-sum support $\mathrm{U}$ with ${i\in \mathrm{U}}$. 
From each of the other servers, the user retrieves $N(N-1)^{k-2}$ ${(k-1)}$-sum symbols supported on ${\mathrm{U}\setminus\{i\}}$.
Therefore, across the other ${N-1}$ servers there are ${N(N-1)^{k-1}}$ available ${(k-1)}$-sum symbols supported on ${\mathrm{U}\setminus\{i\}}$, which matches the number of $k$-sum symbols ${N(N-1)^{k-1}}$ supported on $\mathrm{U}$ that the user retrieves from the fixed server.
Thus, every $k$-sum symbol involving $\mathrm{X}_i$ yields one recoverable subpacket of $\mathrm{X}_i$, and all $N^{f}$ subpackets of $\mathrm{X}_i$ appearing in one server are recoverable.
Consequently, across all $N$ servers, the total number of recoverable subpackets of $\mathrm{X}_i$ is ${N^{f+1}}$ (${=N^{g}}$), which coincides with the subpacketization level ${L^{*} = N^{g}}$ used by the scheme. 
This completes the proof of recoverability for each demand message $\mathrm{X}_i$ with ${i\in \mathrm{S}_2}$.

\subsection{Proof of Privacy} 
We next show that the scheme for the regime ${D\leq K/2}$ satisfies the privacy condition. 
This in turn guarantees privacy for the scheme in the regime ${D>K/2}$:
the first phase does not compromise privacy since the privacy condition is trivially satisfied for the common demand messages, and the second phase invokes the scheme for the regime ${D\leq K/2}$. 

Since the partitioning of messages into sets and blocks is fully determined by $(K,D)$, the user retrieves from each server the same number of singleton symbols and, for every $k$, the same number of $k$-sum symbols for each admissible $k$-message support, independent of the demand index set.
Hence, the scheme yields the same query structure for all candidate demand index sets.
The indices of the demand messages influence only the choice of subpacket indices in the retrieved symbols.
Since the messages are partitioned into subpackets independently and the subpacket indices are assigned via a private random permutation (known only to the user), the indices observed by any server do not indicate which messages belong to the user's demand.
Thus, the servers learn nothing about the user's demand index set, and privacy is guaranteed.

\section{An Illustrative Example}
\label{sec:example_PSSR_cons_dem}
In this section, we present an illustrative example of the proposed PCBR scheme. 

Consider a set of ${K=5}$ messages replicated across ${N=2}$ servers.
We index the messages by $1,2,3,4,5$, and for simplicity, we denote them by $a,b,c,d,e$, respectively. 

Suppose a user wishes to retrieve ${D=2}$ messages, indexed by one of the following ${E=4}$ candidate demand index sets:
\[\mathrm{W}_1 = \{1,2\},\quad \mathrm{W}_2 = \{2,3\},\]
\[\mathrm{W}_3 = \{3,4\},\quad \mathrm{W}_4 = \{4,5\}.\]

In this example, we have 
\[f = \left\lfloor \frac{K}{D}\right\rfloor = 2, \quad g = \left\lceil \frac{K}{D}\right\rceil = 3,\]
and ${M=K - Df = 1}$ since ${D \nmid K}$. 
Thus, ${\mathrm{S}_1=\{1,3,5\}}$ has size ${Mg = 3}$ and is partitioned into ${g=3}$ blocks: 
\[\mathrm{S}_{1,1}=\{1\}, \quad  \mathrm{S}_{1,2}=\{3\}, \quad \mathrm{S}_{1,3}=\{5\},\] each of size ${M=1}$. 
Likewise, ${\mathrm{S}_2=\{2,4\}}$ has size ${(D-M)f=2}$ and is partitioned into ${f=2}$ blocks: 
\[{\mathrm{S}_{2,1}=\{2\}}, \quad \mathrm{S}_{2,2}=\{4\},\] each of size ${D-M=1}$.

Additionally, ${L^{*} = N^{g} = 8}$, so each message is randomly and independently divided into $8$ subpackets, labeled as $a_1, \dots, a_8$, $b_1, \dots, b_8$, $c_1, \dots, c_8$, $d_1, \dots, d_8$, and $e_1, \dots, e_8$.

For this example, the scheme yields the following values of $T_{\mathrm{U}}$ for all non-empty subsets ${\mathrm{U}\subseteq[1:5]}$:
\[
T_{\{1\}}=1,\; T_{\{2\}}=2,\; T_{\{3\}}=1,\; T_{\{4\}}=2,\; T_{\{5\}}=1,
\]
\[
T_{\{1,3\}}=1,\; T_{\{1,5\}}=1,\; T_{\{3,5\}}=1,\; T_{\{2,4\}}=2, 
\]
\[T_{\{1,3,5\}}=1,\]
and $T_{\mathrm{U}}=0$ for all other $\mathrm{U}$. 
For instance, ${T_{\{1\}}=1}$, ${T_{\{2\}}=2}$, and ${T_{\{1,3\}}=1}$ mean that, from each server, the user retrieves one singleton symbol containing a subpacket of message $a$, two singleton symbols each containing a subpacket of message $b$, and one $2$-sum symbol equal to the sum of one subpacket of $a$ and one subpacket of $c$, respectively.

\begin{table}[t]
\renewcommand{\arraystretch}{1.105}
\centering
\caption{Query table for the case $\mathrm{W}=\mathrm{W}_1=\{1,2\}$\\
(demand messages: $a$ and $b$)}
\label{tab:PSSR_cons_dem_W1}
\vspace{0.2em}
\scalebox{1.125}{
\begin{tabular}{|c|c|}
\hline
Server 1 & Server 2\\
\hline\hline
$a_{1}$ & $a_{2}$ \\
$b_{1}$, $b_{2}$ & $b_{3}$, $b_{4}$ \\
$c_{1}$ & $c_{2}$ \\
$d_{1}$, $d_{2}$ & $d_{3}$, $d_{4}$ \\
$e_{1}$ & $e_{2}$ \\
\hline\hline
$a_{3}+c_{2}$ & $a_{4}+c_{1}$\\
$a_{5}+e_{2}$ & $a_{6}+e_{1}$ \\
$c_{3}+e_{3}$ & $c_{4}+e_{4}$ \\
$b_{5}+d_{3}$, $b_{6}+d_{4}$ & $b_{7}+d_{1}$, $b_{8}+d_{2}$ \\
\hline\hline
$a_{7}+c_{4}+e_{4}$ & $a_{8}+c_{3}+e_{3}$ \\
\hline
\end{tabular}
}
\end{table}

\begin{table}[t]
\renewcommand{\arraystretch}{1.105}
\centering
\caption{Query table for the case $\mathrm{W}=\mathrm{W}_2=\{2,3\}$\\
(demand messages: $b$ and $c$)}
\label{tab:PSSR_cons_dem_W2}
\vspace{0.2em}
\scalebox{1.125}{
\begin{tabular}{|c|c|}
\hline
Server 1 & Server 2\\
\hline\hline
$a_{1}$ & $a_{2}$ \\
$b_{1}$, $b_{2}$ & $b_{3}$, $b_{4}$ \\
$c_{1}$ & $c_{2}$ \\
$d_{1}$, $d_{2}$ & $d_{3}$, $d_{4}$ \\
$e_{1}$ & $e_{2}$ \\
\hline\hline
$a_{2}+c_{3}$ & $a_{1}+c_{4}$\\
$a_{3}+e_{3}$ & $a_{4}+e_{4}$ \\
$c_{5}+e_{2}$ & $c_{6}+e_{1}$ \\
$b_{5}+d_{3}$, $b_{6}+d_{4}$ & $b_{7}+d_{1}$, $b_{8}+d_{2}$ \\
\hline\hline
$a_{4}+c_{7}+e_{4}$ & $a_{3}+c_{8}+e_{3}$ \\
\hline
\end{tabular}
}
\end{table}

\begin{table}[t]
\renewcommand{\arraystretch}{1.105}
\centering
\caption{Query table for the case $\mathrm{W}=\mathrm{W}_3=\{3,4\}$\\
(demand messages: $c$ and $d$)}
\label{tab:PSSR_cons_dem_W3}
\vspace{0.2em}
\scalebox{1.125}{
\begin{tabular}{|c|c|}
\hline
Server 1 & Server 2\\
\hline\hline
$a_{1}$ & $a_{2}$ \\
$b_{1}$, $b_{2}$ & $b_{3}$, $b_{4}$ \\
$c_{1}$ & $c_{2}$ \\
$d_{1}$, $d_{2}$ & $d_{3}$, $d_{4}$ \\
$e_{1}$ & $e_{2}$ \\
\hline\hline
$a_{2}+c_{3}$ & $a_{1}+c_{4}$\\
$a_{3}+e_{3}$ & $a_{4}+e_{4}$ \\
$c_{5}+e_{2}$ & $c_{6}+e_{1}$ \\
$b_{3}+d_{5}$, $b_{4}+d_{6}$ & $b_{1}+d_{7}$, $b_{2}+d_{8}$ \\
\hline\hline
$a_{4}+c_{7}+e_{4}$ & $a_{3}+c_{8}+e_{3}$ \\
\hline
\end{tabular}
}
\end{table}

\begin{table}[t]
\renewcommand{\arraystretch}{1.105}
\centering
\caption{Query table for the case $\mathrm{W}=\mathrm{W}_4=\{4,5\}$\\
(demand messages: $d$ and $e$)}
\label{tab:PSSR_cons_dem_W4}
\vspace{0.2em}
\scalebox{1.125}{
\begin{tabular}{|c|c|}
\hline
Server 1 & Server 2\\
\hline\hline
$a_{1}$ & $a_{2}$ \\
$b_{1}$, $b_{2}$ & $b_{3}$, $b_{4}$ \\
$c_{1}$ & $c_{2}$ \\
$d_{1}$, $d_{2}$ & $d_{3}$, $d_{4}$ \\
$e_{1}$ & $e_{2}$ \\
\hline\hline
$a_{3}+c_{3}$ & $a_{4}+c_{4}$\\
$a_{2}+e_{3}$ & $a_{1}+e_{4}$ \\
$c_{2}+e_{5}$ & $c_{1}+e_{6}$ \\
$b_{3}+d_{5}$, $b_{4}+d_{6}$ & $b_{1}+d_{7}$, $b_{2}+d_{8}$ \\
\hline\hline
$a_{4}+c_{4}+e_{7}$ & $a_{3}+c_{3}+e_{8}$ \\
\hline
\end{tabular}
}
\end{table}
 
Tables~\ref{tab:PSSR_cons_dem_W1},~\ref{tab:PSSR_cons_dem_W2},~\ref{tab:PSSR_cons_dem_W3}, and~\ref{tab:PSSR_cons_dem_W4} list the queries for the two servers when the user's demand index set is $\mathrm{W}_1$, $\mathrm{W}_2$, $\mathrm{W}_3$, and $\mathrm{W}_4$, respectively.
Below, Tables~\ref{tab:PSSR_cons_dem_W1} and~\ref{tab:PSSR_cons_dem_W2} are discussed in detail, with the subpacket indexing explained together with the corresponding recovery process to establish correctness. 
The proofs for Tables~\ref{tab:PSSR_cons_dem_W3} and~\ref{tab:PSSR_cons_dem_W4} follow in the same way.

Consider Table~\ref{tab:PSSR_cons_dem_W1}, which corresponds to the demand index set $\mathrm{W}_1 = \{1,2\}$, i.e., the demand messages are $a$ and $b$. 
Assigning distinct subpacket indices to the retrieved singleton symbols of each message from each server, the user retrieves $a_1$, $b_1$, $b_2$, $c_1$, $d_1$, $d_2$, and $e_1$ from Server~1 and $a_2$, $b_3$, $b_4$, $c_2$, $d_3$, $d_4$, and $e_2$ from Server~2.
Thus, the user directly recovers $a_1$, $a_2$, $b_1$, $b_2$, $b_3$, and $b_4$.

Next, consider $2$-sum symbols.
In the symbol involving messages $c$ and $e$, since both $c$ and $e$ are interference messages, they must be assigned subpacket indices that have not been previously used at any server.
Thus, they are assigned index $3$ in the symbol retrieved from Server~1 and index $4$ in the symbol retrieved from Server~2.

In the symbol involving $a$ and $c$ retrieved from Server~1, the demand message $a$ is assigned index $3$ that has not been previously used at any server, while $c$ is assigned index $2$ since $c_2$ was previously recovered from Server~2. 
Similarly, in the symbol retrieved from Server~2, $a$ and $c$ are assigned indices $4$ and $1$, respectively.
Thus, the user recovers $a_3$ and $a_4$.

In the symbol involving $a$ and $e$ retrieved from Server~1, $a$ is assigned index $5$ that has not been previously used at any server, while $e$ is assigned index $2$ since $e_2$ was previously recovered from Server~2. 
Likewise, in the symbol retrieved from Server~2, $a$ and $e$ are assigned indices $6$ and $1$, respectively.
Thus, the user recovers $a_5$ and $a_6$.

In the two symbols involving $b$ and $d$ retrieved from Server~1, the demand message $b$ is assigned indices $5$ and $6$ that have not been previously used at any server, while $d$ is assigned indices $3$ and $4$ since $d_3$ and $d_4$ were previously recovered from Server~2. 
Likewise, $b$ is assigned indices $7$ and $8$ and $d$ is assigned indices $1$ and $2$ in the two symbols retrieved from Server~2. 
Thus, the user recovers $b_5$, $b_6$, $b_7$, and $b_8$.

Finally, consider the $3$-sum symbol involving $a$, $c$, and $e$ retrieved from Server~1. 
The demand message $a$ is assigned index $7$ that has not been previously used at any server, while $c$ and $e$ are assigned index $4$ since the $2$-sum $c_4+e_4$ was previously retrieved from Server~2. 
Likewise, $a$, $c$, and $e$ are assigned indices $8$, $3$, and $3$ in the symbol retrieved from Server~2. 
Thus, the user recovers $a_7$ and $a_8$. 
This completes the subpacket indexing and the recovery process, which establishes correctness for the case ${\mathrm{W} = \mathrm{W}_1}$. 

We now consider Table~\ref{tab:PSSR_cons_dem_W2}, which corresponds to the demand index set $\mathrm{W}_2 = \{2,3\}$, i.e., the demand messages are $b$ and $c$.
Subpacket indexing of the retrieved singleton symbols and $2$-sum symbols involving $b$ and $d$ follows the same process as in the case $\mathrm{W}=\mathrm{W}_1$. 
Thus, the user recovers $b_1, \dots, b_8$, as well as $c_1$ and $c_2$.

In the $2$-sum symbol involving the interference messages $a$ and $e$ retrieved from Server~1, $a$ and $e$ are assigned index $3$ that has not been previously used at any server.
Similarly, they are assigned index $4$ in the symbol retrieved from Server~2.

In the symbol involving $a$ and $c$ retrieved from Server~1, $c$ is assigned index $3$ that has not been previously used at any server, while $a$ is assigned index $2$ since $a_2$ was previously recovered from Server~2. 
Likewise, in the symbol retrieved from Server~2, $a$ and $c$ are assigned indices $1$ and $4$, respectively.
Thus, the user recovers $c_3$ and $c_4$. 

In the symbol involving $c$ and $e$ retrieved from Server~1, $c$ is assigned index $5$ that has not been previously used at any server, while $e$ is assigned index $2$ since $e_2$ was previously recovered from Server~2.
Likewise, in the symbol retrieved from Server~2, $c$ and $e$ are assigned indices $6$ and $1$, respectively.
Thus, the user recovers $c_5$ and $c_6$.

Finally, in the $3$-sum symbol involving $a$, $c$, and $e$ retrieved from Server~1, $c$ is assigned index $7$ that has not been previously used at any server, while $a$ and $e$ are assigned index $4$ since the $2$-sum $a_4+e_4$ was previously retrieved from Server~2.
Likewise, $a$, $c$, and $e$ are assigned indices $3$, $8$, and $3$ in the symbol retrieved from Server~2. 
Thus, the user recovers $c_7$ and $c_8$. 
This completes the subpacket indexing and the recovery process, which establishes correctness for the case ${\mathrm{W} = \mathrm{W}_2}$.

User privacy follows immediately because the query structure is identical for every candidate demand index set, the message subpackets are indexed independently and uniformly at random across messages, and within each server’s query no subpacket index is used more than once.

In this example, the user retrieves $13$ symbols from each server, each symbol having the size of one message subpacket, while the user's demand consists of two messages, each partitioned into $8$ subpackets. 
The resulting rate is therefore
\[
\frac{2\times 8}{2\times13}=\frac{8}{13}, 
\] 
which matches the rate $R$ in~\eqref{eq:PSSR_cap_cons_dem}. 
Moreover, the scheme requires subpacketization level ${L^{*} = 8}$, matching the lower bound $L_{*}$ in~\eqref{eq:PCBR_lb_L} for this instance.

Finally, we compare the performance of our PCBR scheme with that of MPIR schemes for the same parameters ${N=2}$, ${K=5}$, and ${D=2}$. 
Since the PCBR demand family in this example consists of the four contiguous pairs and is therefore a restriction of the full MPIR demand space, any MPIR scheme for these parameters can be directly applied here. 
Nevertheless, the best-known MPIR scheme for this setting, due to~\cite{HWS2025}, achieves rate $82/135$ and requires subpacketization level $82$. 
In contrast, our PCBR scheme achieves the higher rate $8/13$ while operating with a smaller subpacketization level $8$.

\balance

\bibliographystyle{IEEEtran}
\bibliography{PIR_PC_Refs}

\end{document}